\documentstyle[amssymb,prb,twocolumn, aps]{revtex}

\input epsf
\begin{document}

\smallskip 
\twocolumn[\hsize\textwidth\columnwidth\hsize\csname@twocolumnfalse\endcsname

\title{Allowed and forbidden transitions in artificial hydrogen and helium atoms}
\author{Toshimasa Fujisawa$^{\ast }$, David Guy Austing$^{\ast ,\dagger }$,
Yasuhiro Tokura$^{\ast }$, Yoshiro Hirayama$^{\ast ,\ddagger }$ \& Seigo
Tarucha$^{\ast ,\S ,\Vert }$}
\address{$^{\ast }$NTT Basic Research Laboratories, NTT Corporation,
3-1 Morinosato-Wakamiya, Atsugi, 243-0198, Japan\\
$^{\dagger }$Institute for Microstructual Sciences M23A, National Research
Council of Canada, Ottawa, Ontario K1A 0R6, Canada\\
$^{\ddagger }$CREST, 4-1-8 Honmachi, Kawaguchi, 331-0012, Japan\\
$^{\S }$University of Tokyo, Bunkyo-ku, Tokyo, 113-0033, Japan\\
$^{\Vert }$ERATO Mesoscopic Correlation Project, 3-1 Morinosato-Wakamiya,
Atsugi, 243-0198, Japan
\newline
}
\maketitle

] \narrowtext

\newpage

{\bf The strength of radiative transitions in atoms is governed by selection
rules \cite{Bethe}. Spectroscopic studies of allowed transitions in hydrogen
and helium provided crucial evidence for the Bohr's model of an atom.
``Forbidden'' transitions, which are actually allowed by higher-order
processes or other mechanisms, indicate how well the quantum numbers
describe the system. We apply these tests to the quantum states in
semiconductor quantum dots (QDs), which are regarded as artificial atoms.
Electrons in a QD occupy quantized states in the same manner as electrons in
real atoms \cite{QDotReview,AshooriAA,TaruchaAA,LPKexcitationAA}. However,
unlike real atoms, the confinement potential of the QD is anisotropic, and
the electrons can easily couple with {\it phonons} of the material\cite%
{FujisawaScience}. Understanding the selection rules for such QDs is an
important issue for the manipulation of quantum states. Here we investigate
allowed and forbidden transitions for phonon emission in one- and
two-electron QDs (artificial hydrogen and helium atoms) by electrical
pump-and-probe experiments, and find that the total spin is an excellent
quantum number in artificial atoms. This is attractive for potential
applications to spin based information storage.}

The QD we study is located in a circular pillar (diameter of 0.5 $\mu $m)
fabricated from an AlGaAs/InGaAs heterostructure (see Figs. 1a and 1b) \cite%
{TaruchaAA}. Electrons can be injected from, and collected by, the source
(s) and drain (d) electrodes through asymmetric tunneling barriers, whose
tunneling rates are $\Gamma _{s}\sim $ (3 ns)$^{-1}$ and $\Gamma _{d}\sim $
(100 ns)$^{-1}$ for the barrier nearest to the source and the drain,
respectively. The number of electrons in the dot, $N$, can be controlled
exactly by applying a voltage, $V_{g}$, to the surrounding gate electrode
(g). Electrons are confined in an In$_{0.05}$Ga$_{0.95}$As quantum well
(thickness $a$ = 12 nm) in the vertical ($z$) direction, and by a
two-dimensional harmonic potential in lateral ($x$ and $y$) direction
appropriate for small $N$. The first few electrons occupy the 1s and 2p
orbitals associated with the lateral confinement. Since our QD does not have
circular symmetry \cite{TokuraElliptic,Matagne}, orbital degeneracy is
lifted even at zero magnetic field, and only a two-fold spin degeneracy is
expected. For this non-circular (approximately elliptic) QD, we still use
1s, 2p, ... to label the orbitals\ for convenience. This non-circularity
does not affect much our discussion or the underlying physics. The following
experiments are carried out at a temperature, $T$, of $\sim $100 mK, unless
otherwise stated.

\begin{figure}[tbp]
\epsfxsize=3.3in \epsfbox{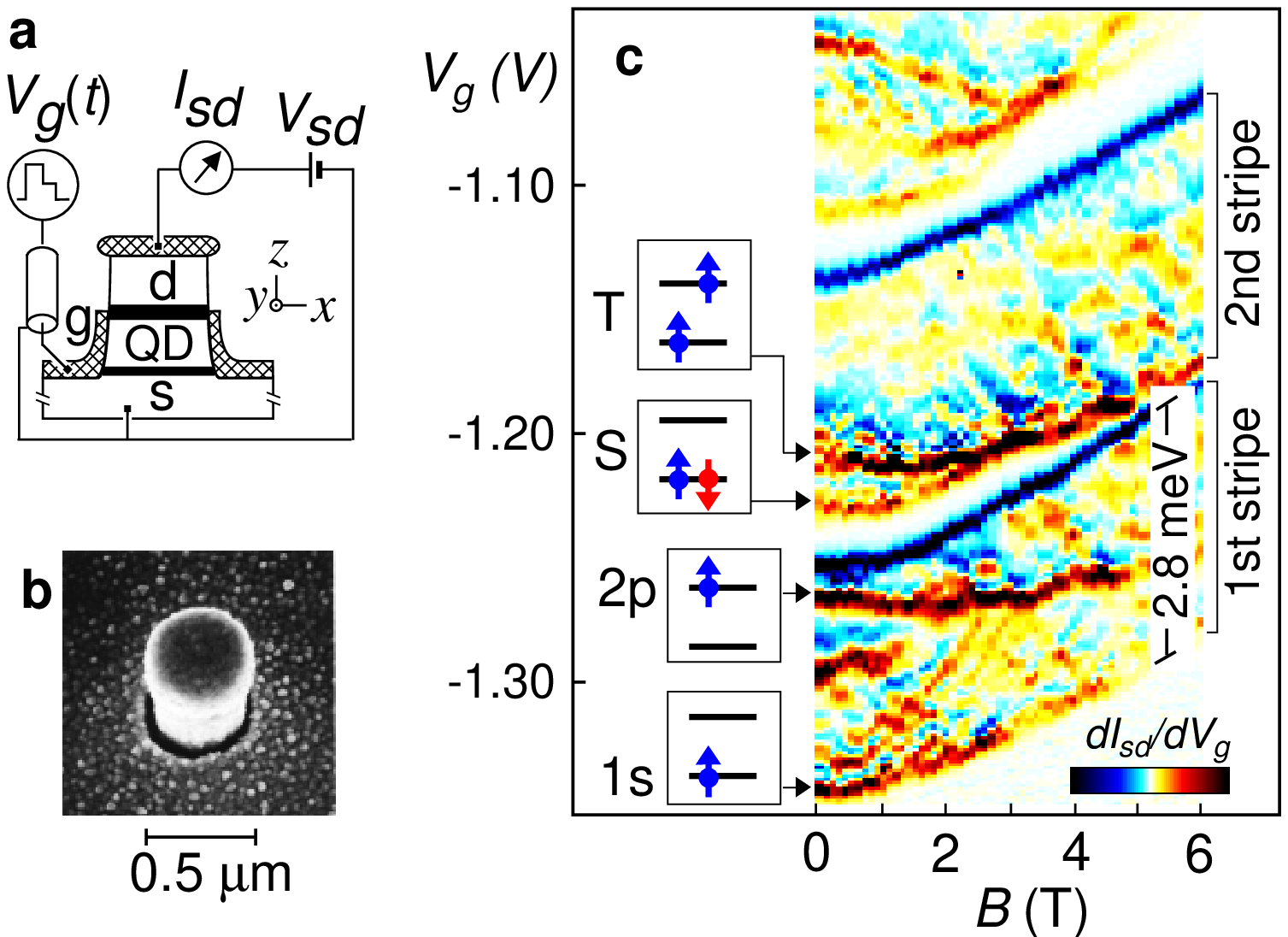}
\caption{Artificial hydrogen and helium atoms. {\bf a}, Schematic set
up for the pulse measurement on the vertical quantum dot. The In$_{0.05}$Ga$%
_{0.95}$As quantum dot (QD) is connected to the source (s) and the drain (d)
electrodes made of Si doped GaAs by asymmetric Al$_{0.22}$Ga$_{0.78}$As
tunneling barriers (7 nm thick for the lower one and 8.5 nm for the upper
one. Their tunneling rates, $\Gamma _{s}\sim $ (3 ns)$^{-1}$ and $\Gamma
_{d}\sim $ (100 ns)$^{-1}$, are obtained by separate measurements). The gate
electrode (g) is connected to a pulse generator, which produces a gate
voltage, $V_{g}(t)$, of rectangular or double-step shape. The measurements
are performed in a dilution refrigerator in a magnetic field $B$ = 0 - 5 T
applied parallel to the $z$ direction. {\bf b}, A
scanning-electron-micrograph of a control device. {\bf c}, $dI_{sd}/dV_{g}$
for the $N$ = 1 and 2 QD taken with a large source-drain voltage ($V_{sd}=$
2.8 mV). The first (second) current stripe gives information about the $N=$
1 (2) QD. The peaks indicated by the arrows show the $B$-field evolution of
the ground state (lowest edge of each stripe) and the first excited state.
The relevant 1- and 2-electron configurations are also shown, in which the
lower (upper) horizontal line represents 1s (2p) orbital.}
\end{figure}

First we investigate the $N=$ 1 QD (artificial hydrogen), in which single
electron occupies the 1s orbital (the ground state) or the 2p orbital (the
first excited state). The energy spectrum of these states can be obtained by
tunneling spectroscopy \cite{LPKexcitationAA}. The current, $I_{sd}$,
increases stepwise with increasing $V_{g}$ (giving peaks in $dI_{sd}/dV_{g}$%
) each time an empty dot state enters the transport window. Color plot of $%
dI_{sd}/dV_{g}$ vs $V_{g}$ traces taken as a function of magnetic field ($B$%
) applied in the $z$ direction are shown in Fig. 1c. The peak spacings
within a given stripe can be related to the energy spacings between
corresponding states. The energy spacing between the 1s and 2p states, $%
\varepsilon _{1s-2p}$, of the $N$ = 1 QD deduced from the first current
stripe is plotted in Fig. 2e. The lateral confinement of our QD can be
approximated by two orthogonal harmonic potentials in the $x$ and $y$ plane %
\cite{TokuraElliptic}. The characteristic confinement energies, $\hbar
\omega _{x}\sim $ 2.5 meV and $\hbar \omega _{y}\sim $ 5.5 meV ($\hbar
\equiv h/2\pi $ is the Planck's constant), are obtained by fitting $%
\varepsilon _{1s-2p}$ (see solid line in Fig. 2e). The Zeeman splitting of
the electron spin is small ($E_{Z}\sim $ 0.1 meV at $B=$ 5 T), and is
unresolved in our experiments.

We now focus on the energy relaxation from the 2p state to the 1s state in
the $N=1$ QD, which changes the electron's orbital momentum but preserves
the spin. Electrical pump-and-probe experiments are performed by applying a
gate voltage of rectangular-shaped time dependence, $V_{g}(t)$, which
switches between $V_{l}$ and $V_{h}$ as shown in Fig. 2a, to the gate.
Experimental details are given in Refs. 9 \& 10
. First, the $N$ = 0 QD is prepared during the low-phase of the pulse ($%
V_{g}=V_{l}$, see Fig. 2b). The period, $t_{l}$ = 100 - 200 ns, is made
sufficiently long to ensure that both the 1s and 2p states are empty. When
the pulse is switched on ($V_{g}=V_{h}$, see Fig. 2c) such that only the 2p
state is located in the transport window, an electron can be injected into
the 2p state from the source (pump) with a time constant, $\Gamma
_{s}^{-1}\sim $ 3 ns. The electron can only escape to the drain (probe) more
slowly, with a time constant, $\Gamma _{d}^{-1}\sim $ 100 ns. However, this
escape process can be interrupted by the relaxation into the 1s ground
state. Thus, the current contains information about the relaxation lifetime, 
$\tau _{1s-2p}$. We measure the averaged dc current, $I_{p}$, under the
application of the pulse train. Figure 2d shows how this current changes
with the pulse length, $t_{h}$. $I_{p}$ is then converted into an average
number of tunneling electrons per pulse, $\langle n_{t}\rangle
=I_{p}(t_{h}+t_{l})/e$ ($e$ is the elementary charge). From a detailed
analysis of the rate equations including all possible tunneling processes
for the $N=1$ QD, we find $\langle n_{t}\rangle \sim \Gamma _{d}\tau
_{1s-2p}[1-\exp (-t_{h}/\tau _{1s-2p})]$ under the condition, $\Gamma
_{s}^{-1}\lesssim \tau _{1s-2p}<\Gamma _{d}^{-1}$, required for the
relaxation time measurement \cite{PRBpulse}. We made sure that this
condition is satisfied in all measurements for the $N=1$ QD. The relaxation
time thus estimated from the rise time of $\langle n_{t}\rangle $ is $\tau
_{1s-2p}\sim $ 10 ns for the case in Fig. 2d. The small saturation value of $%
\langle n_{t}\rangle \sim $ 0.02 indicates very efficient relaxation.

We discuss the origin of the relaxation in artificial hydrogen. For an
energy spacing in the meV region, and at low temperature (thermal energy $%
\sim $10 $\mu $eV $\ll \varepsilon _{1s-2p}$), spontaneous emission of a 
{\it phonon}, rather than a photon, dominates the relaxation process \cite%
{FujisawaScience}. The energy spacing coincides with the acoustic-phonon
energy in the linear dispersion regime. Because of the discrete energy of
the states, the relaxation involves emission of a phonon of energy equal to $%
\varepsilon _{1s-2p}$ (with corresponding phonon wavelength, $\lambda
_{1s-2p}$). In our experiment, $\varepsilon _{1s-2p}$, and hence $\lambda
_{1s-2p}$, vary with $B$ as shown in Figs. 2e and 2f. Note also that the
characteristic sizes $l_{x}$ and $l_{y}$ for the lateral dimensions of the
QD decrease with increasing $B$. The strength of the electron-phonon
interaction is expected to be suppressed for $\lambda _{1s-2p}$ smaller than
the characteristics size of the QD (phonon bottleneck effect \cite{Benisty}%
). Therefore, the $B$ dependence of $\tau _{1s-2p}$ in Fig. 2g is because
the phonon emission is suppressed (i.e. $\tau _{1s-2p}$ increases) with
decreasing $B$ when $\lambda _{1s-2p}$ becomes shorter than $a$, $l_{x}$,
and $l_{y}$.

\begin{figure}[tbp]
\epsfxsize=3.3in \epsfbox{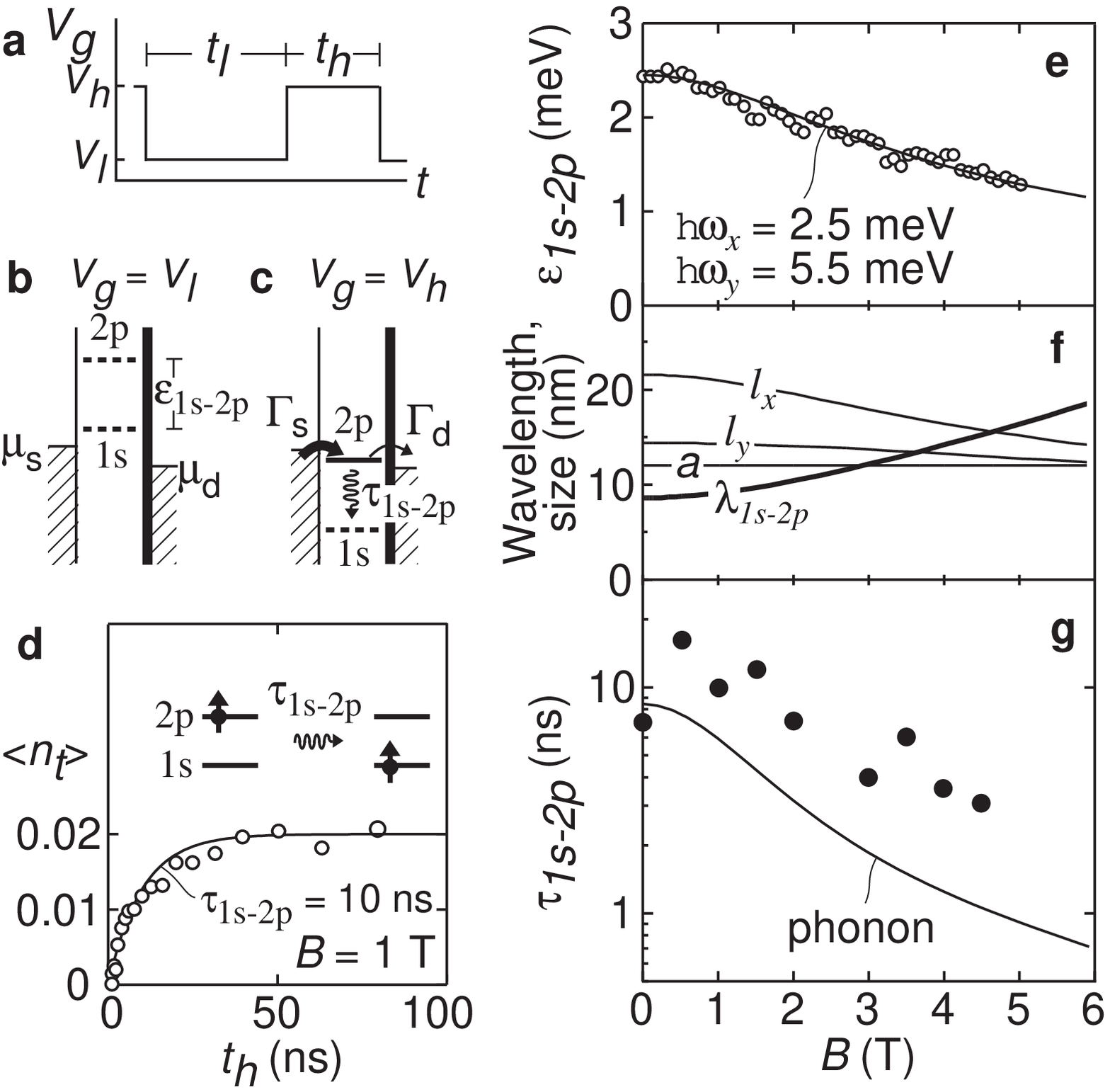}
\caption{ Relaxation time of a one-electron QD (artificial hydrogen
atom). {\bf a}, Schematic of pulse waveform used for the electrical
pump-and-probe experiment.\ {\bf b }and {\bf c}, Schematic energy diagrams
along the z direction showing low and high pulse situations. The thick and
thin vertical lines denote the asymmetric tunneling barriers. States in the
electrodes are filled up to the Fermi energies, $\mu _{s}$ for the source
and $\mu _{d}$ for the drain. The source-drain voltage, $V_{sd}$, opens a
small transport window $eV_{sd}=\mu _{s}-\mu _{d}\sim $ 0.1 meV. Solid and
dashed horizontal lines denote filled and empty single-particle states,
respectively. When $V_{g}=V_{l}$ {\bf b}, the 1s and 2p states are located
above $\mu _{s}$ and $\mu _{d}$. When $V_{g}=V_{h}$ {\bf c}, only the 2p
state is located in the transport window. The 2p state is pumped from the
source at a tunneling rate, $\Gamma _{s}\sim $ (3 ns)$^{-1}$, and probed at
a slower rate, $\Gamma _{d}\sim $ (100 ns)$^{-1}$. The current measures the
momentum relaxation time of the 2p state, $\tau _{1s-2p}$.\ {\bf d}, The
average number of tunneling electrons per pulse, $\langle n_{t}\rangle $,
measured at 1 T. The relaxation time, $\tau _{1s-2p}$ = 10 ns, is obtained
from the exponential curve (solid line) fitted to the data. The inset shows
the electron configuration before and after relaxation. {\bf e through g},
Magnetic field, $B$, dependence of {\bf e} the energy spacing between the 2p
excited state and the 1s ground state, $\varepsilon _{1s-2p}$, {\bf f} the
longitudinal acoustic phonon wavelength, $\lambda $, and characteristic
sizes of the QD ($a$, $l_{x}$ and $l_{y}$), and {\bf g} the energy
relaxation time, $\tau _{1s-2p}$. In {\bf f}, $\lambda $ is calculated for
the phonon at energy $\varepsilon _{1s-2p}$ using a GaAs sound velocity of
5100 m/s. The characteristic lateral size in the $x$/$y$ direction is given
by $l_{x/y}=\sqrt{\hbar /m^{\ast }}(\omega _{x/y}^{2}+\omega
_{c}^{2}/4)^{-1/4}$, where $m^{\ast }$ is the effective mass. The solid line
in {\bf g} is calculated for spontaneous emission of an acoustic phonon.}
\end{figure}

In order to be quantitative, we calculate the phonon emission rate from
Fermi's golden rule including both deformation and piezoelectric types of
coupling with standard GaAs material parameters \cite{Seeger,Bockelmann}.
For simplicity, the calculation is done for a circular dot, whose effective
confinement energy is $\hbar \omega _{eff}=\hbar \sqrt{\omega _{x}\omega
_{y}(1+\omega _{c}^{2}/(\omega _{x}+\omega _{y})^{2})}$, where $\omega _{c}$
is the cyclotron frequency. This assumption is reasonable because Coulomb
interactions in an elliptic QD just scale with $\hbar \omega _{eff}$ \cite%
{TokuraElliptic}. As shown by the solid line in Fig. 2g, we find good
agreement with the data. The difference by about a factor of 2 or 3 might
come from the assumptions about the confinement potential and uncertainties
in the material parameters. Thus, the fast energy relaxation in $N=$ 1 QD
can be understood by spontaneous emission of a phonon. It should be noted
that no clear selection rule for orbital momentum is expected, because the
electron-phonon interaction cannot be approximated by a dipole interaction,
and because of the anisotropic confinement potential.

In contrast, the relaxation time is remarkably different for a $N=$ 2 QD
(artificial helium). At low magnetic fields (see second stripe in Fig. 1c
for $B$ 
\mbox{$<$}
2.5 T), the many-body ground state is a spin-singlet (labeled S) with two
antiparallel-spin electrons occupying the 1s orbital, while the first
excited state is a spin-triplet (labeled T) with two parallel-spin
electrons, one each occupying the 1s and 2p orbitals \cite%
{TaruchaAA,LPKexcitationAA}. Because of direct Coulomb and exchange
interactions, the energy spacing between the two states, $\varepsilon _{S-T}$
($\sim $0.6 meV at $B=$ 0 T), is smaller than $\varepsilon _{1s-2p}$. Energy
relaxation from the first excited state (T) to the ground state (S) not only
involves the same change in orbital momentum as that in the $N=$ 1 QD, but
also requires a spin-flip because of Pauli exclusion (see inset of Fig. 3e).
A simple phonon-emission transition from the spin-triplet to the
spin-singlet is forbidden by spin conservation.

We now investigate to what degree this transition is ``forbidden''. The
simple rectangular pulse technique employed for the $N=$ 1 QD is not useful
for this $N=$ 2 QD transition, because the relaxation lifetime, $\tau _{S-T}$%
, is always beyond the measurable range ($\tau _{S-T}>\Gamma _{d}^{-1}\sim $
100 ns) \cite{PRBpulse}. Instead, we subject the QD to a double-step
voltage-pulse, in which $V_{g}$ is switched between three voltages, $V_{l}$, 
$V_{h}$\ and $V_{m}$ as shown in Fig. 3a.\ First, when $V_{g}=V_{l}$ (Fig.
3b), the $N=$ 1 QD is prepared during a sufficiently long period, $t_{l}$ =
100 ns. When $V_{g}$ is suddenly increased to $V_{h}$ (Fig. 3c), an electron
can enter to create the $N=$ 2 triplet state within the interval $\sim
\Gamma _{s}^{-1}=$ 3 - 7 ns. $V_{g}=V_{h}$ for the duration $t_{h}=$ 100 ns
- 100 $\mu $s, which is much longer than $\Gamma _{s}^{-1}$. The triplet
state may suffer a relaxation process during this time. When $V_{g}$ is
changed to $V_{m}$ (Fig. 3d), an electron in the triplet state can tunnel
out to the drain, if the triplet state is not yet relaxed to the singlet
state. We set the period, $t_{m}$ = 300 ns, to be longer than $\Gamma
_{d}^{-1}$ so we can read out the signal. We repeatedly apply the
double-step pulse (effectively $\sim $10$^{7}$ times) to obtain a reliable
current $I_{p}$, and evaluate the average number of tunneling electrons, $%
\langle n_{t}\rangle =I_{p}(t_{l}+t_{h}+t_{m})/e$. Since the current
measures the unrelaxed electron number, $\langle n_{t}\rangle =A\exp
(-t_{h}/\tau _{S-T})$ for the condition $\Gamma _{s}^{-1}\lesssim \tau _{S-T}
$ (no upper limit in principle), where $A\sim 1$ is approximately the ratio
of $\Gamma _{s}^{-1}$ for the triplet to that for the singlet. Figure 3e
shows a typical measurement of $\langle n_{t}\rangle $ at 0 T, indicating a
relaxation time of $\tau _{S-T}\sim $ 200 $\mu $s. This relaxation time is 4
to 5 orders of magnitude longer than that observed in the $N=$ 1 QD.

\begin{figure}[tbp]
\epsfxsize=3.3in \epsfbox{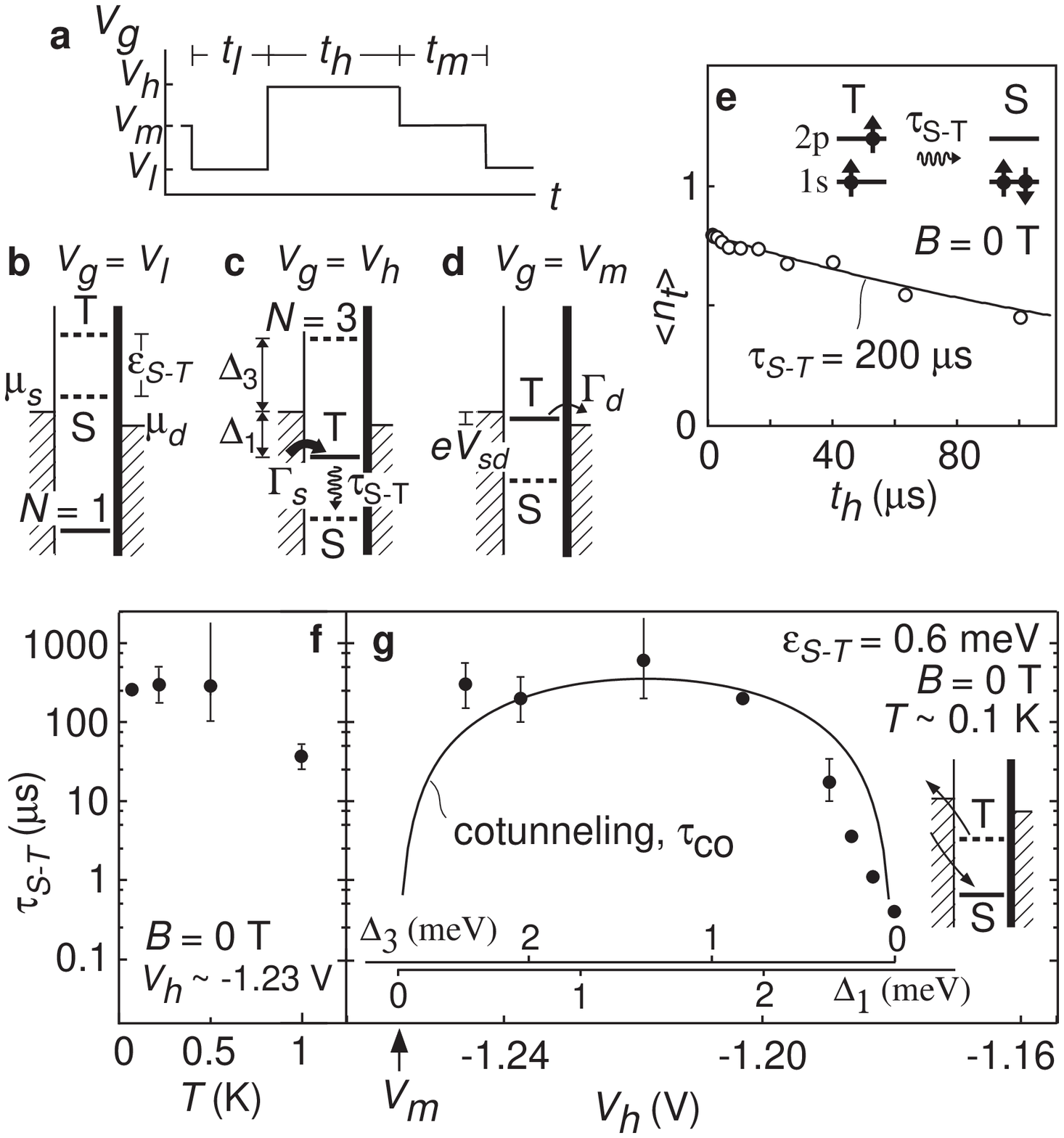}
\caption{Relaxation time of a two-electron QD (artificial helium
atom). {\bf a}, Schematic of double-step pulse waveform to measure extremely
long relaxation times. {\bf b through d}, Schematic energy diagram showing
low, high and intermediate pulse situations. Solid and dashed horizontal
lines denote filled and empty many-body states, respectively. $\mu _{s}-\mu
_{d}\sim $ 0.1 meV. When $V_{g}=V_{l}$ {\bf b}, the spin-singlet ground
state S and the spin-triplet first excited state T are located above $\mu
_{s}$ and $\mu _{d}$. The system will always become the $N$ = 1 QD after a
sufficiently long period, $t_{l}$ = 100 ns. When $V_{g}=V_{h}$ {\bf c}, the
QD can be excited to the triplet state within $\Gamma _{s}^{-1}\sim $ 7 ns.
The triplet state can then relax to the singlet state during the period, $%
t_{h}$ = 0.1 - 100 $\mu $s. When $V_{g}=V_{m}$ {\bf d}, the triplet state is
probed by allowing an electron to tunnel into the drain. This period is
fixed at $t_{m}$ = 300 ns. {\bf e}, Average number of tunneling electrons
per pulse, $\langle n_{t}\rangle $ at 0 T. The relaxation time, $\tau _{S-T}$
= 200 $\mu $s, is obtained from the exponential decay (solid line). The
inset shows the electron configuration before and after relaxation. {\bf f},
Temperature, $T$, dependence of the relaxation time $\tau _{S-T}$ at 0 T. 
{\bf g}, The gate voltage, $V_{h}$, dependence of $\tau _{S-T}$. $V_{h}$ is
also converted into $\Delta _{1}$ and $\Delta _{3}$ energy scales. $\Delta
_{1}$ and $\Delta _{3}$ are indicated in {\bf c}. The solid line is
calculated for cotunneling processes. The inset shows a schematic of these
inelastic cotunneling processes.}
\end{figure}

We find no clear $B$ dependence of $\tau _{S-T}$ (always longer than 100 $%
\mu $s), at least for the energy spacings between 0.6 meV at $B=$\ 0 T and
0.24 meV at $B=$ 2 T (not shown). We also investigate the temperature
dependence of $\tau _{S-T}$ (Fig. 3f). No clear change is observed up to 0.5
K. $\tau _{S-T}$ decreases above 0.5 K, where thermal excitation from the QD
to the electrodes becomes important.

On the other hand, we do find that $\tau _{S-T}$ strongly depends on the
high gate-pulse voltage $V_{h}$, during which relaxation takes place, as
shown in Fig. 3g. Although $V_{h}$ is swept deep into the $N=$ 2 Coulomb
blockade region (-1.27 V $<V_{g}<$ -1.16 V), $\tau _{S-T}$ decreases rapidly
at $V_{h}\sim $ -1.18 V. This $V_{h}$ dependence implies a strong influence
of the source and drain electrodes. Even though the Coulomb blockade is
robust in the suppression of transport, higher-order tunneling processes can
contribute to the relaxation and change $\tau _{S-T}$. An electron in the
dot can be replaced with an electron of opposite spin from the electrodes
(see inset of Fig. 3g). This results in energy loss in the QD, whereas the
electrode gains the same energy. This inelastic cotunneling rate, $\tau
_{co}^{-1}$, is estimated by considering second-order tunneling processes %
\cite{Cotunneling,MEto,Sukhorukov}. For the relaxation mechanisms considered
here, the $N=$ 2 QD can relax virtually through $N=$ 1 or $N=$ 3
intermediate states. Note that this process does not cause a net current
even at a finite voltage of $e|V_{sd}|<\varepsilon _{S-T}$ \cite%
{DeFranceschi}. Assuming $V_{sd}=$ 0 V and zero temperature for simplicity,
we obtain $\tau _{co}^{-1}=\varepsilon _{S-T}(\hbar \Gamma _{s}+\hbar \Gamma
_{d})^{2}(\Delta _{1}^{-1}+\Delta _{3}^{-1})^{2}/h$. Here, $\Delta _{1}$ and 
$\Delta _{3}$, respectively are the energies required to excite the initial $%
N=$ 2 triplet state to the $N=$ 1 and 3 intermediate states, as indicated in
Fig. 3c. We can extract $\Delta _{1}$ and $\Delta _{3}$ from $V_{h}$, and
the values are shown in Fig. 3g. The solid line shows $\tau _{co}$, the
relaxation time due to cotunneling, calculated with experimentally deduced
parameters ($\varepsilon _{S-T}=$ 0.6 meV, and $(\Gamma _{s}+\Gamma
_{d})^{-1}=$ 7 ns). Clearly the observed relaxation time can be well
understood by inelastic cotunneling.

Our observations for $N=$ 1 and 2 QDs can be compared with real atoms \cite%
{Bethe}. The transition from the 2p state to the 1s state in atomic hydrogen
is allowed by photon emission (the Lyman $\alpha $ transition line), while
that in artificial hydrogen is allowed by {\it phonon} emission. The
transition from the spin-triplet state to the spin-singlet state is
forbidden by conservation of the total spin for both atomic helium and
artificial helium. Moreover, these ``forbidden'' transitions can occur by
collisions with electrons for the helium atom, and by cotunneling for the $%
N= $ 2 QD. Spin conservation also applies to any transition in light atoms,
and probably likewise to few-electron QDs. The difference between the
allowed and forbidden transitions leads to more than 11 orders of magnitude
difference in the relaxation times for real hydrogen and helium atoms. Our
observation of 4 to 5 orders of magnitude difference in our $N=1$ and 2
artificial atoms is not as high, but is still surprisingly large. Note that
this difference would become larger if cotunneling in QDs can be suppressed
by using thicker tunneling barriers. Very importantly, the large difference
between $\tau _{1s-2p}$ and $\tau _{S-T}$ originates from the fact that
other effects, such as spin-orbit and hyperfine interactions \cite%
{OpticalOrientation,Khaetskii}, must have only a weak effect on the breaking
of the ``forbidden'' symmetries. We now discuss how small these hidden
contributions are by focusing on the spin-orbit interactions.

Spin-orbit interactions are predicted to give the dominant contribution to
spin relaxation in GaAs QD systems \cite{Khaetskii}, although this is still
an extremely small effect. For simplicity, we consider the spin-orbit
interaction energy, $\Delta _{so}$, only for coupling between the 1s and 2p
orbitals, but including all effects which mix spin and orbital degrees of
freedom. Simple perturbation theory\cite{WPHalperin} predicts that the
relaxation time from the triplet to the singlet is given by $\tau
_{S-T,so}\sim (\varepsilon _{S-T}/\Delta _{so})^{2}\tau
_{phonon}(\varepsilon _{S-T})$. Here, $\tau _{phonon}^{-1}(\varepsilon
_{S-T})$ is the phonon emission rate at the phonon energy, $\varepsilon
_{S-T}$, and we know that $\tau _{phonon}$ is well accounted for by the
electron-phonon interaction. Therefore, we can deduce an upper bound of $%
\Delta _{so}<$ 4 $\mu $eV from our observations ($\tau _{S-T}>$ 200 $\mu $%
s). This value is close to the spin-orbit induced spin splitting energy ($%
\sim $ 2.5 $\mu $eV) observed in a GaAs two-dimensional electron gas system %
\cite{BychkovRashba}. {}Note that spin-orbit interactions are significantly
enhanced in nanoparticles, e.g. copper, probably because of impurities or
interfaces \cite{WPHalperin}, but our vertical semiconductor QD are largely
free of these undesirable factors.

Our experiments indicate that the spin degree of freedom in QDs is well
isolated from the orbital degree of freedom. This is particularly attractive
for applications to spin memories and spin quantum bits (qubits) \cite%
{Loss,Recher}. For a simple scheme involving just a single-electron spin in
a magnetic field, the spin-orbit interactions can degrade the energy
relaxation time ($T_{1}$) of a spin qubit. We estimate the dominant
contribution, $T_{1,so}$, using a perturbative approach, $T_{1,so}\sim
(\varepsilon _{1s-2p}/\Delta _{so})^{2}\tau _{phonon}(\varepsilon _{Z})$.
Since $\Delta _{so}<$ 4 $\mu $eV, this yields $T_{1,so}>$ 1 ms for a Zeeman
splitting $\varepsilon _{Z}\sim $ 0.1 meV and $\varepsilon _{1s-2p}\sim $
1.2 meV at $B$ $\sim $ 5 T ($T_{1,so}>$ 100 $\mu $s at $B$ $\sim $ 9 T).
This $T_{1,so}$ is thus comparable to that obtained by
electron-spin-resonance for donor states in GaAs \cite{Seck}, and is much
longer than the time required for typical one- and two-qubit operations \cite%
{GuptaAwschalom}. Note that small spin-orbit interactions are also desirable
with respect to the dephasing time ($T_{2}$) of a spin qubit\cite{Kavokin}.
Our results therefore encourage further research in the use of the spin
degree of freedom in QDs.


\bigskip

{\bf Acknowledgement}

We thank G. E. W. Bauer, T. Honda, T. Inoshita, A. V. Khaetskii, L. P.
Kouwenhoven for discussions and help.

\vspace{0.2 cm}

Correspondence and requests for materials should be addressed to T.F.
(e-mail: fujisawa@will.brl.ntt.co.jp).

\end{document}